# Recombination of Geminate (OH,e$_{aq}^-$) Pairs in Concentrated Alkaline Solutions: Lack of Evidence For Hydroxyl Radical Deprotonation. [1]

Rui Lian, Robert A. Crowell, [*] Ilya A. Shkrob, David M. Bartels, [a)]

Dmitri A. Oulianov and David Gosztola

*Chemistry Division, Argonne National Laboratory, Argonne, IL 60439*

**Abstract**

Picosecond dynamics of hydrated electrons and hydroxyl radicals generated in 200 nm photodissociation of aqueous hydroxide and 400 nm (3-photon) ionization of water in concentrated alkaline solutions were obtained. No deprotonation of hydroxyl radicals was observed on sub-nanosecond time scale, even in 1-10 M KOH solutions. This result is completely at odds with the kinetic data for deprotonation of OH radical in dilute alkaline solutions. We suggest that the deprotonation of hydroxyl radical is slowed down dramatically in concentrated alkaline solutions.

---

[*] To whom correspondence should be addressed: *Tel* 630-252-8089, *FAX* 630-2524993, *e-mail:* rob_crowell@anl.gov.

a) Current address: Radiation Laboratory, University of Notre Dame, Notre Dame, Indiana 46556; *e-mail:* bartels@hertz.rad.nd.edu.

## 1. Introduction

In recent years, there has been renewed interest in theoretical and experimental studies of the dynamics of electron detachment from simple inorganic anions in liquids, such as halides in water (e.g., [1]) and sodide in THF (e.g., [2]). One of the less studied of these photoreactions is the dissociation of aqueous hydroxide. This photoreaction yields a geminate pair of hydroxyl radical (OH) and hydrated electron ($e_{aq}^-$)

$$OH^- \xrightarrow{h\nu} OH + e_{aq}^- \tag{1}$$

with a quantum efficiency of 0.11 [3,4]. This geminate pair is of fundamental importance for the radiation- and photo- chemistry of water, because the same species, OH and $e_{aq}^-$, are formed in the ionization of neat water [5], where the backward electron transfer reaction

$$OH + e_{aq}^- \longrightarrow OH^- \qquad\qquad k_2 = 3 \times 10^{10}\ M^{-1}\ s^{-1}\ [6] \tag{2}$$

accounts for 80% of geminate electron decay (the remaining 20% of the photoelectrons decay is via reaction with a hydronium ion) [7]. Although rxn. (2) is one of the fastest reactions of hydrated electrons, in alkaline solutions its rate is comparable to that of the deprotonation of OH radical

$$OH + OH^- \rightleftharpoons O_{aq}^- + H_2O \tag{3}$$

(The $pK_a$ of the hydroxyl radical at 25 °C is 11.54 to 12.1 [8-11]). Most estimates for the rate constant of forward rxn. (3) cluster around $(1.2\text{-}1.3) \times 10^{10}\ M^{-1}\ s^{-1}$ (e.g., [12]), though Hickel, Cortfizen, and Sehested [13] gave $(6.3 \pm 1.3) \times 10^9\ M^{-1}\ s^{-1}$ for the forward and $10^6\ s^{-1}$ for the backward reaction (previous estimates for the latter were as high as $(9.2\text{-}9.6) \times 10^7\ s^{-1}$ [12,14]). All of these kinetic data were obtained in dilute alkaline solutions.

Rxn. (3) plays the central role in the radiation chemistry of strongly alkaline solutions, such as high-level radioactive waste stored within the US-DOE complex. Note that the neutralization of the hydronium ion formed in the radiation- or photo- induced water ionization,

$$H_2O \xrightarrow{h\nu, \gamma} OH + e^-_{aq} + H_3O^+ \qquad (4)$$

is very rapid in these concentrated alkaline solutions,

$$H_3O^+ + OH^- \longrightarrow 2\ H_2O \qquad k=1.1\times10^{11}\ M^{-1}\ s^{-1}\ [15] \qquad (5)$$

so the three-particle spur formed in rxn. (4) is rapidly (in a matter of tens of picoseconds) converted to a pair formed in rxn. (1). Thus, the geminate $(OH, e^-_{aq})$ pair generated via rxn. (1) makes a useful reference system for studying charge separation dynamics in radiolytically induced rxn. (4).

In this work, we study the effect of alkalinity on the geminate recombination dynamics of $(OH, e^-_{aq})$ pairs generated in rxn. (1), by single photon 200 nm photoexcitation of hydroxide, and rxn. (4), by 3-photon 400 nm ionization of water in these alkaline solutions. In 1-10 M KOH solutions, a change in the recombination dynamics of the electron due to the occurrence of rapid (forward) rxn. (3) was expected. Contrary to these expectations, Crowell et al. [16] observed no concentration dependence for 1-10 M hydroxide. This result is doubly surprising, because in addition to rapid rxn. (3) converting OH to $O^-$, it is known from other studies [17] that the survival probability of geminate pairs generated by electron photodetachment from aqueous anions rapidly decreases with the ionic strength of the solution. This effect is thought to originate through the cation association with the electron and its anion precursor [17]. For hydroxide anion photoexcited at 193 nm, ca. 6% decrease in the free electron yield per 1 M ionic strength (changed by addition of NaClO$_4$) has been observed by Sauer et al. [17] Yet the addition of 1-10 M of KOH [16] and 1-9 M NaClO$_4$ [17] had negligible effect on the electron dynamics within the first 600 ps after the photoexcitation pulse. Cation association, increased bulk viscosity, and ionic atmosphere drag on the electron mobility all seem to have very little, if any, effect on sub-nanosecond dynamics of the $(OH, e^-_{aq})$ pairs.

A possible rationale for our failure to observe the OH deprotonation indirectly, through its effect on the electron dynamics, would be that these dynamics are not too sensitive to the replacement of OH by O$^-$. Indeed, it is known that in dilute alkaline solutions, the rate constants of the bulk reaction

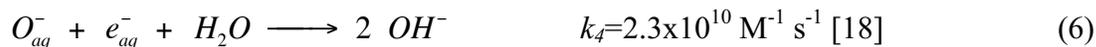

$$O^-_{aq} + e^-_{aq} + H_2O \longrightarrow 2\ OH^- \qquad k_4=2.3\times10^{10}\ M^{-1}\ s^{-1}\ [18] \qquad (6)$$

and rxn. (2) are fairly similar, and these two constants may become even more similar in the presence of atmospheric ions screening the repulsive Coulomb potential of the O$^-$ anion. Below, it is demonstrated that this rationale does not account for our kinetic observations, because the anticipated disappearance of OH by rxn. (3) was not observed even when the OH radical, rather than its geminate partner, was detected.

**2. Experimental.**

The kinetic measurements reported below were obtained using a 1 kHz Ti:sapphire chirped pulse amplification setup (see refs. [19] for more detail) which yielded Gaussian, 60 fs FWHM, 3 mJ pulses of 800 nm light. These pulses were used to generate 400 nm (second harmonic) or 200 nm (fourth harmonic) pump pulses. Up to 20 µJ of the 200 nm light (300-350 fs FWHM pulse) and 200 µJ of 400 nm light (150-200 fs FWHM pulse) was produced by harmonic generation. The 800 nm and 266 nm (third harmonic) probe pulses were derived from the same beam. See refs. [16] and [19] for more detail. The pump and probe beams were perpendicularly polarized and focused to round spots of 60-150 µm and 20-30 µm in radius, respectively. The maximum fluences of the 200 nm and 400 nm photons were 0.025 and 0.4 J/cm$^2$, respectively. The vertical bars shown in the kinetics (Fig. 2) represent 95% confidence limits for each data point. Typically, 150-200 points acquired on a quasi-logarithmic grid out to 600 ps were used to obtain these kinetics. A short-lived "spike" was observed within the duration of 200 nm or 400 nm pump pulse, which is due to simultaneous absorption of this pump pulse and 266 nm probe pulse (i. e., "1+1" nonlinear absorbance) [20]. This "spike" is not shown in the kinetic plots given below. In very concentrated alkali solutions, fewer 200 nm photons are available for this "1+1" process due to the increased sample absorption, and the "spike" is less prominent.

Hydroxide solutions of low concentration were prepared by mixing 1 M KOH volumetric standard (Aldrich) with deaerated nanopure water; concentrated alkaline solutions were prepared by dissolving solid KOH pellets (Aldrich, highest purity) in this standard solution. The laser beams were crossed at 6.5° at the surface of a 150 μm thick jet. To prevent the absorption of $CO_2$ from the air, this high speed jet was enclosed in a box that was constantly purged with dry $N_2$. An all 316 stainless steel and Teflon flow system was used to pump the solution through the jet.

### 3. 266 nm absorptivity of OH, O$^-$, and $e_{aq}^-$.

Both $e_{aq}^-$ and OH absorb at 266 nm, but the molar absorptivities for these two species in the UV are not known accurately. For the OH radical, the overall shape of the absorption band is known [21-26], but the estimates for the absolute extinction coefficient vary by 30%. For $e_{aq}^-$, the wavelength of 266 nm corresponds to a narrow intervalley between the *1s → CB* absorption band in the VIS and near UV and a strong bound-to-bound transition at 180 nm that involves O *2p* orbitals of water molecules [23]. Different absorption spectra and estimates for the absorption coefficient have been given for this region in the literature [23,25]. Perhaps, the most consistent estimates were given by Boyle et al. [23] (who used UV flash photolysis of water): $\varepsilon_{266}(OH)=370$ M$^{-1}$ cm$^{-1}$ and $\varepsilon_{266}(e_{aq}^-)=720$ M$^{-1}$ cm$^{-1}$ and by Nielsen et al. [25] (who used pulse radiolysis of neutral aqueous solutions): $\varepsilon_{266}(OH)=420$ M$^{-1}$ cm$^{-1}$ and $\varepsilon_{266}(e_{aq}^-)=600$ M$^{-1}$ cm$^{-1}$. In concentrated salt solutions (including alkaline solutions), the vis band of the hydrated electron shifts to the blue [17,27,28]; the amount of this effect on the UV absorbance of the electron is not known.

The spectrum of O$^-$ is known [24,29], as this species can be generated by pulse radiolysis of $N_2O$-saturated alkaline solutions. Above 300 nm, the UV spectra of O$^-$ (in 1 M NaOH) and OH (in *pH*=7 solution) are nearly identical. Below 300 nm, the absorbance of the OH radical passes through a maximum at 230 nm (see Fig. 3 in ref. [24]) whereas the spectrum of O$^-$ slowly increases towards the UV. While the exact shape of the spectrum at *λ*<220 nm (where hydroxide absorbs strongly) was not reproducible, both groups that studied the O$^-$ spectrum obtained $\varepsilon_{266}(O^-)=220$ M$^{-1}$ cm$^{-1}$ (vs. 460 M$^{-1}$ cm$^{-1}$ for OH) [24]. In fact, 266 nm is the most convenient wavelength to observe the progress of rxn. (3) spectroscopically since the ratio of the extinction coefficients of the OH and O$^-$ radicals reaches a maximum of ca. 2.1 at this wavelength.

Using the estimates given above, it is easy to demonstrate that the conversion of the $(OH, e_{aq}^-)$ pair to the $(O^-, e_{aq}^-)$ pair via rxn. (3) should decrease the total absorbance at 266 nm by at least 23%. Such a change would be readily observable in the kinetic traces, provided that rxn. (3) occurred on the expected time scale. That condition requires [OH$^-$] > 1 M. Note that backward rxn. (3) is negligible on our time scale.

**4. Results.**

*a. 200 nm pump - 266 nm probe kinetics.*

Fig. 1 shows 200 nm pump - 266 nm probe kinetics obtained in 0.1, 1, 3, 5, and 10 M solution of KOH. The absorbance signal scaled linearly with the pump power; 2-photon excitation of water by 200 nm light was negligible due to high absorptivity of OH$^-$ at 200 nm ($\varepsilon_{200}$(OH$^-$)=1000 M$^{-1}$ cm$^{-1}$). To facilitate the comparison, the kinetics shown in Fig. 1 were normalized at 10 ps. As can be seen from this plot, all of these normalized 266 nm kinetics are the same; furthermore, these kinetics are identical to the 800 nm kinetics obtained for the same photosystem [16]. Only the hydrated electron absorbs at the latter wavelength. The anticipated loss of the 266 nm absorbance due to rapid forward rxn. (3) was not observed.

Confronted with this result, two control experiments were carried out in order to establish (i) whether the OH radicals can be observed using our setup and (ii) whether the rapid deprotonation of these OH radicals in other photosystems can be observed. Answering the second question is important because it has not been fully established whether rxn. (1) is the *only* reaction of photoexcited OH$^-$. In particular, rapid deprotonation of a vibrationally hot OH radical can occur in concert with the electron detachment, yielding the O$^-$ radical directly:

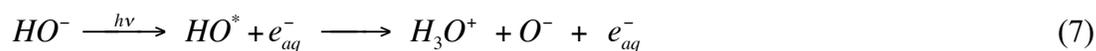

$$HO^- \xrightarrow{h\nu} HO^* + e_{aq}^- \longrightarrow H_3O^+ + O^- + e_{aq}^- \qquad (7)$$

In such a case, the expected transformation of OH to O$^-$ via rxn. (3) would not occur because no OH radicals are present in the photolysate in the first place. We remind the reader that a concerted proton and electron transfer (written as rxn. (1)) similar to rxn. (7) is considered as the most likely mechanism for low-energy ionization of liquid water (see ref. [5] and references therein). Conceivably, a similar photoprocess might have occurred in the laser excitation of OH$^-$.

To answer the first question, we studied 200 nm photodissociation of aqueous hydrogen peroxide [19]. Transient absorption kinetics of the OH radicals at 266 nm were observed in photolysis of 1 M $H_2O_2$ (see ref. [19] for more detail). The OH radicals rapidly escaped from the solvent cage into the bulk with a time constant of 30 ps. Using the estimates for OH absorptivity given in section 3, the quantum yield of peroxide decomposition at 500 ps was 0.44 [19] vs. 0.49±0.07 [30] or 0.47±0.03 [31] at 254 nm and 0.45±0.06 at 222 nm [32] obtained by others. In addition to showing that OH radicals can be observed using our setup, the molar absorptivity estimates given in section 3 were reaffirmed in this control experiment.

Unfortunately, 200 nm photodissociation of $H_2O_2$ cannot be used to answer whether it is possible to directly observe the OH deprotonation on the fast time scale. To observe rxn. (3) within the first 600 ps (which is the time window of our study), the concentration of OH$^-$ must be high, whereas $pK_a$ for $H_2O_2$ and OH are similar, 11.7 and 11.8, respectively [15]. Consequently, $H_2O_2$ is fully deprotonated in the alkaline solutions; the photoexcitation of $HO_2^-$ yields OH and O$^-$ radicals. The resulting OH radical rapidly reacts with $HO_2^-$ yielding $O_2^-$ (with rate constant of 7x10$^9$ M$^{-1}$ s$^{-1}$ [15]) that strongly absorbs at 266 nm [33], and this reaction competes with rxn. (3). Thus, the only way to observe the deprotonation is to have large excess of OH$^-$ with respect to $HO_2^-$. However, in such a case, most of the 200 nm photons would be absorbed by hydroxide as it has 5 times higher absorbance at 200 nm than $HO_2^-$ [23,24]. Furthermore, $HO_2^-$ is thermodynamically unstable and rapidly dissociates in a high-speed jet. Consequently, a different strategy was used to generate OH radicals in the strongly alkaline solutions.

*b. 400 nm photon pump - 266 nm probe kinetics.*

For 400 nm light, 2-photon absorptivity of 1 M hydroxide is 4 cm/TW vs. 3-photon absorptivity of 270 $cm^2/TW^3$ for neat water [16]. Therefore, when the 400 nm irradiance exceeds 0.1-2 $TW/cm^2$, most of the electrons generated in the 400 nm photoexcitation of 1 M KOH solution originate through 3-photon ionization of water. This photoionization yields $H_3O^+$ ions (neutralized in 10 ps via rxn. (5)) and hydroxyl radical. By comparing 800 nm (electron) and 266 nm (electron + hydroxyl) kinetics obtained under identical excitation conditions it is possible to deduce whether the expected extra loss of the 266 nm absorbance due to ongoing forward rxn. (3) occurs.

Fig. 2 demonstrates a comparison between the 266 nm and 800 nm kinetics obtained by 400 nm excitation of 1 M KOH. Since the solution is relatively transparent to the 400 nm light, a weak second "spike" (<3% of the prompt "spike") was observed due to "1+1" nonlinear absorbance that involved the pump pulse reflected by the back surface of the jet. This second "spike" is not shown in the plot. The transient kinetics were normalized at 5-10 ps, at which time this nonlinear absorbance becomes negligible. Under the excitation conditions of Fig. 2, ca. 75% of the electrons were formed via 3-photon ionization of water. At the higher pump power, the 266 nm kinetics become noisier, due to thermal lensing in the jet. It appears that within the confidence limits of our kinetic measurement, the two kinetic traces at 266 and 800 nm are identical. No deprotonation of the OH radical was observed. Thus, postulating rxn (7) to account for our observations for hydroxide served no purpose. Advocating the use of the Occam principle, we thereby conclude that rxn. (1) rather than rxn. (7) is the main dissociative path for one-photon excitation of $OH^-$.

**5. Discussion.**

To summarize, the expected consequences of rxn. (3) were not observed in strongly alkaline solutions on the sub-nanosecond time scale, neither in electron photodetachment from hydroxide (section 4.a) nor in water photoionization (section 4.b).

Our failure to observe rxn. (3) directly, at 266 nm, or indirectly, at 800 nm, leaves only two possible interpretations:

The first is that rxn. (3) is at least 10 times slower than estimated from the extrapolation of kinetic data obtained from dilute alkaline solutions (see the Introduction). The second is that the extinction coefficients given in the literature for the species involved (section 3) are grossly incorrect. For OH and $e_{aq}^-$, these coefficients and the UV spectra have been measured at neutral *pH*, whereas our kinetic measurements were carried out in 1-10 M alkaline solutions. If the UV spectrum for the OH radical exhibits a concentration dependent band shift, that may render the difference between the absorbances of the $(OH, e_{aq}^-)$ pair and the $(O^-, e_{aq}^-)$ pair too small to be observed in our pump-probe experiments. Furthermore, if the blue shift of the electron spectrum in the alkaline solutions is extended to the UV, the molar absorptivity of the electron at 266 nm may increase, and that decreases the relative reduction in the 266 nm absorbance when the $(OH, e_{aq}^-)$ pair is converted to $(O^-, e_{aq}^-)$ pair via rxn. (3).

We believe that the second rationale is less likely. The 230 nm band of OH radical originates through an intramolecular transition in this radical; it is unlikely that this band demonstrates significant ionic strength dependence. The blue shift in the absorption spectrum of $O^-$ is likely (see below), but this shift would only increase the reduction in the 266 nm absorbance. The electron spectrum might shift to the blue in the UV, but the expected shift (judged from the vis band shift) is ca. 1 nm per 1 M ionic strength [17,35], and the absorption spectrum around 266 nm is flat. Our sensitivity was such that even a 5-10% reduction in the 266 nm absorbance due to rxn. (3) would have been observed at least on some time scale, and it seems very unlikely that an increase in the 266 nm absorbance of the electron due to the band shift can account for our observations.

Thus, by exclusion of other options, we conclude that the rate of forward rxn. (3) is dramatically reduced in concentrated (1-10 M) alkaline solutions. To the best of our knowledge, such a claim is neither supported nor contradicted by the existing body of pulse radiolysis and flash photolysis studies of such solutions.

The cause for the retardation of OH deprotonation in concentrated alkali is currently unclear. Many aqueous anions, such as halides and pseudohalides, exhibit the so-called CTTS (charge transfer to solvent) bands in their UV spectra; the red "tail" of the $O^-$ spectrum strikingly resembles those CTTS bands. It is known that these CTTS bands shift to the blue due to the constriction of the solvation cage around the anion [27]; it is likely that such a constriction also occurs for $OH^-$ and $O^-$, changing the energetics of rxn. (3). Furthermore, slowing down of forward rxn. (3) may be the trivial consequence of an increase in the solution viscosity, from 1 cP in water to 3.4 cP in 10 M KOH. While the data on molecular and ionic diffusion coefficients in concentrated alkaline solutions are very scarce, it is known that for dioxygen [34] and ferro- and ferri- cyanide [35] and formate [36] anions the product of the diffusion coefficient and the bulk viscosity is constant in 1-4 M KOH solutions. If this holds for more concentrated alkali solutions, rxn. (3) may slow down because the diffusion of OH and $OH^-$ slows down in these viscous solutions. Interestingly, we do not observe such an effect for migration of solvated electron since the geminate recombination dynamics for $(OH, e^-_{aq})$ pairs do not change with KOH concentration. This curious lack of the ionic strength effect for some reactions of hydrated electron, e.g. $e^-_{hyd} + e^-_{hyd} \longrightarrow 2\ OH^- + H_2$ reaction, has already been reported by one of us [37].

## 5. Acknowledgement.

This work was performed under the auspices of the Office of Science, Division of Chemical Science, US-DOE under contract number W-31-109-ENG-38. The authors thank Prof. S. E. Bradforth of USC, Dr. S. V. Lymar of BNL, and Drs. C. D. Jonah and M. C. Sauer, Jr. of ANL for useful discussions.

**Figure Captions.**

**Fig. 1.**

Normalized kinetics for 266 nm absorbance observed in 200 nm excitation of KOH solutions. The molar concentrations used are given in the plot (two series of such measurements are given together; the detection scheme that was used differed between these two series). The maximum $\Delta OD_{266}$ signal (after the "1+1" spike) was $(2-5) \times 10^{-3}$ OD. No effect of alkaline concentration on the time profile of these kinetics is observed, despite the fact that rxn. (3) is expected to be occurring.

**Fig. 2.**

Transient absorbances at 800 nm (to the left; solid line) and 266 nm (to the right; filled circles with 95% confidence bars) observed in 400 nm excitation of 1 M KOH. At 800 nm, the absorption signal is from the hydrated electron, at 266 nm – from the hydrated electron and OH radical. The latter is expected to convert to $O^-$ in 80 ps via rxn. (3), resulting in 23% loss in the total absorbance (section 3). This loss was not observed.

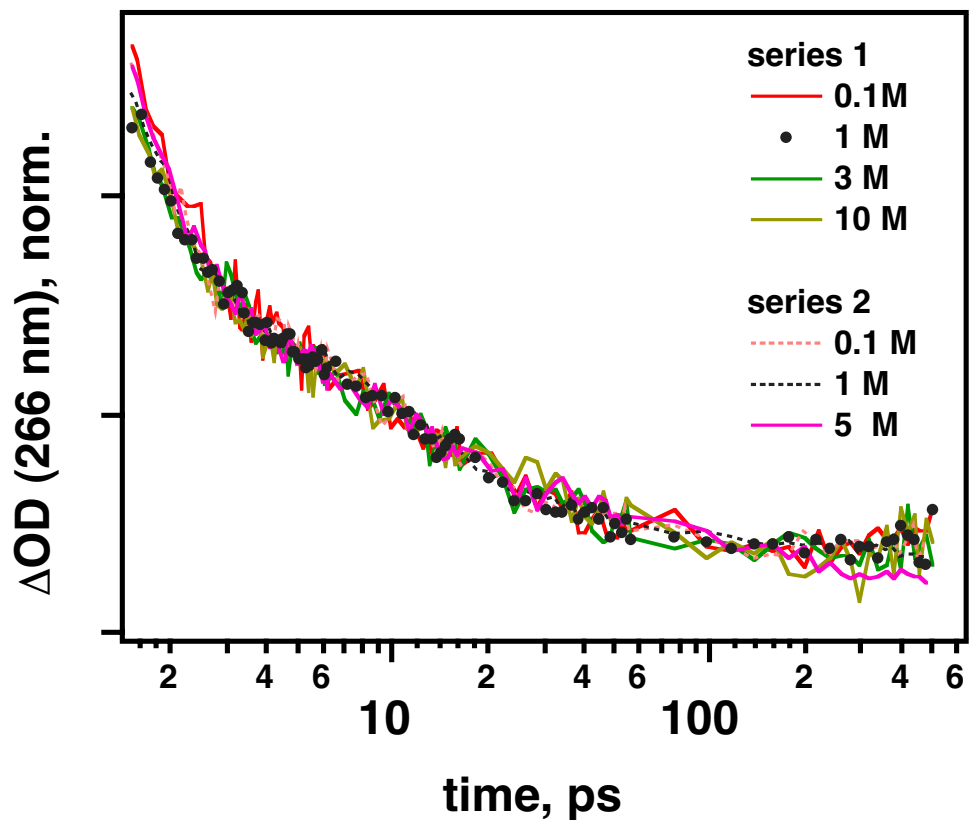

**Figure 1.**

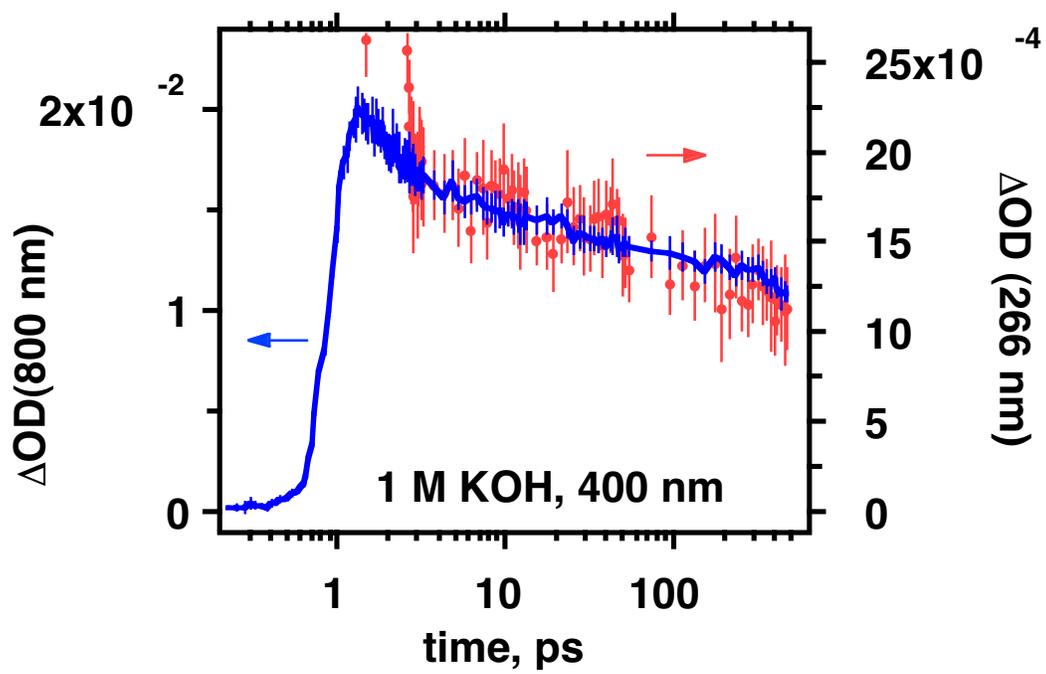

**Figure 2.**